\documentclass[11pt,fleqn,twoside]{article}
\usepackage{graphicx,cite,xurl,pdfpages}
 \usepackage{fancyhdr}
\date{}
\newlength{\hsbw}
\newcommand{\memo}[1]{\mbox{}\par\vspace{0.25in}
\setlength{\hsbw}{\linewidth}
\addtolength{\hsbw}{-2\fboxsep}
\addtolength{\hsbw}{-2\fboxrule}
\noindent\fbox{\parbox{\hsbw}{{\bf Memo: }#1}}\vspace{0.25in}}
\renewcommand{\memo}[1]{}
\usepackage[bookmarks=true,hyperfigures=true,colorlinks=true,linkcolor=blue,citecolor=blue,pdfpagemode=fullscreen,plainpages=false,pdfpagelabels]{hyperref}
\usepackage[all]{hypcap}

\newcommand{\arxiv}[1]{\href{https://arxiv.org/abs/#1}{\tt arXiv:#1}}
\title{\vspace*{-1ex}Where AI Assurance Might Go Wrong\\
\normalsize Initial lessons from engineering of critical systems}
\author{Robin E.  Bloomfield (City St George's, University of London
UK)\\and John Rushby (SRI International, Menlo Park CA USA)\\[1ex]
\texttt{r.e.bloomfield@city.ac.uk} \hspace*{2ex} \texttt{rushby@csl.sri.com}}
\begin{document}
\maketitle

\begin{abstract}

We draw on our experience working on system and software assurance and
evaluation for systems important to society to summarise how safety
engineering is performed in traditional critical systems, such as
aircraft flight control.  We analyse how this critical systems
perspective might support the development and implementation of AI
Safety Frameworks.  We present the analysis in terms of: system
engineering, safety and risk analysis, and decision analysis and
support.

We consider four key questions: What is the system?  How good does it
have to be?  What is the impact of criticality on system development?
and How much should we trust it?  We identify topics worthy of further
discussion. In particular, we are concerned that system boundaries are
not broad enough, that the tolerability and nature of the risks are
not sufficiently elaborated, and that the assurance methods lack
theories that would allow behaviours to be adequately assured.

We advocate the use of assurance cases based on Assurance 2.0 to
support decision making in which the criticality of the decision as
well as the criticality of the system are evaluated.  We point out the
orders of magnitude difference in confidence needed in critical rather
than everyday systems and how everyday techniques do not scale in
rigour.

Finally we map our findings in detail to two of the questions posed by
the FAISC organisers and we note that the engineering of critical
systems has evolved through open and diverse discussion.  We hope that
topics identified here will support the post-FAISC dialogues.

\end{abstract}
\thispagestyle{fancy}
\newpage

\tableofcontents
\listoffigures
\newpage

\section{Introduction}\label{introduction}

There are three main ways in which AI Assurance might go wrong: one is
that it addresses the wrong risks, second is that its techniques are
inadequate for the risks it does address, and finally it may fail
to communicate its claims effectively to the public and other
stakeholders.

We are neither AI researchers nor developers, although we are users;
we are outsiders who work on general system and software assurance
(and evaluation), particularly for systems important to society.  We
were among the first to propose structured safety cases
\cite{Bloomfield&Bishop:SSS10} (see \cite[Section 2]{Rushby:Cases15}
for some history) and experience has led us to develop a more rigorous
and critical interpretation that we call Assurance 2.0
\cite{Bloomfield&Rushby:Assurance2,Bloomfield&Rushby24:CBJ,Varadarajan-all:DASC24}
(see the \hyperref[appendix]{Appendix} for an overview
and the collection of papers at \cite{Assurance2-home} for more
details).  However, we have long been interested in the topic of AI
assurance and wrote on the topic more than 35 years ago
\cite{Rushby88:SQA,Bloomfield&Ehrenberger88}, and also recently
\cite{Bloomfield&Rushby:AI24}.  This paper draws on our experience
with critical systems and also with systems using AI.

In Sections \ref{eng} and \ref{analysis} we present a description and
analysis of these topics from experience of engineering critical
systems: learning from success and failure, and what are still hard
problems.  In Section \ref{evolution} we then relate them to two
specific questions posed in the conference Call for Submissions.

\begin{enumerate}
\item

Improving existing safety frameworks: How can existing safety
frameworks be strengthened? How can we adapt best practices from other
industries?

\item

Building on safety frameworks (Seoul Commitment V): How will safety
frameworks need to change over time as AI systems' capabilities
improve?  How do they need to change when AI systems become capable of
posing intolerable levels of risk?
\end{enumerate}

The engineering of critical systems is of course a rich and varied topic
and we have been selective in this analysis, making a judgment of what
we think might be fruitful at this stage in the development and
implementation of AI safety frameworks.

At this point we should digress a little into terminology: in
particular, the AI community uses ``safety'' as an umbrella term to
address a wide range of harms.  In this paper we use the terms safety
and security in the classical sense, as used in critical systems:
safety can be seen as the possibly harmful impact of the system on its
environment and security as harmful impact of the environment on the
system.  We note that general caution is needed in terminology: in
addition to safety and security, hazard, confidence, risk, model and
several other terms all have different and specific meanings in
different communities.

\section{Engineering Critical Systems}
\label{eng}

To set the context, we summarise how safety engineering is performed
in traditional systems, such as aircraft flight control.  This is the
\emph{dependability perspective} \cite{Evidence07,Laprie:terminology}
and we discuss its application to AI in a recent report
\cite{Bloomfield&Rushby:AI24}.  Below, we enumerate (simplified) steps
of the traditional process.

\begin{enumerate}
\item

There is a system context, or world model, generally called the
\textbf{Environment}.  For a flight system such as an autopilot, this
might refer to or describe the overall airplane, aerodynamics,
aircraft behavior, structures, weather, sensors with their failure
modes and frequencies, and so on, including other systems and people
that the system interacts with.  Note that many critical systems are
operated by trained pilots/operators and are not engineered to be used
by the untrained public.

\item

There are \textbf{System Requirements} that describe what the system
(e.g., an autopilot) is to do, principally in terms of the effects it
is to have on the environment (e.g., in {\sc althld} mode it maintains
airplane altitude within some range).  Note, requirements
describe what is to be done, not how to do it.

\item

There is the process of \textbf{Hazard Analysis} that identifies
circumstances in the conjunction of the environment and the system
requirements that have the potential to lead to harm (e.q.,
uncommanded pitch down).  In the nuclear industry a ``design basis''
is defined as the worst-case environmental challenges to be addressed
by the system \cite{ONR-Generic-Assessment} and the financial sector
also uses the concept of ``design basis threats.''

\item
There are safety specifications (generally called \textbf{Safety
Requirements}) that describe constraints on the system requirements and
on acceptable failure rates (e.g., ``not anticipated to occur in the
entire lifetime of all aircraft of one type'') and that also identify
requirements for defence in depth (for example, no single fault may
precipitate a ``catastrophic failure condition'' in a civil aircraft).

\item

There is a process of \textbf{Requirements Validation} that seeks to
establish that the safety requirements eliminate or mitigate all the
hazards.  As the system specification develops, new hazards may be
introduced (e.g., to mitigate a fire in the electronics bay, we add a
fire suppression system and must then consider new hazards associated
with failure of that system).  The whole process iterates until it
stabilizes.  The iteration will assess the extent of defence in depth
needed and the role of other subsystems and people.

\item

There is a \textbf{System Specification} that describes how the
system is to work.  While developing the system specification (or, later,
its implementation) the engineers may encounter circumstances that could
be hazardous.  These are elevated to the systems engineers who may concur
and add new hazards and safety requirements, causing requirements
validation to be revisited.

\item

There is an implementation, and a \textbf{Verification Process} (or
verifier) that establishes that the system specification and its
implementation satisfy the safety requirements (and, with lesser
assurance, the general system requirements).  The verifier may be a
static process (e.g., formal verification of the implementation), or an
active one (i.e., runtime verification using monitors or guards).  There
will be requirements that vary with criticality for different levels of
independence (technical, organisational) of the verification and
development processes.

\item

There is an overall safety or \textbf{Assurance Case} that establishes
that the specific realization of the steps above are sufficient to
provide indefeasible assurance for the top safety claim.  Our
preferred approach to assurance cases in terms of claims, evidence,
and argument, is described in \cite{Bloomfield&Rushby24:CBJ}.  The
assurance case provides an instrument for wider challenge, and support
for decision making on whether to grant authority to operate (e.g.
put an airplane into commercial service).

\end{enumerate}

There is extensive historical experience underlying the development of
this approach, and much data.  In particular there have been \emph{no}
accidents of modern aircraft due to failures of Step 7 (Verification).
This despite the fact that the processes used are quite elementary
(mainly code reviews, static analysis and testing, with very little
formal verification): these are slow and expensive but industry knows
how to do them.  \emph{All} modern aircraft failures have been
attributed to Step 5 (Requirements Validation), which may in turn
implicate Steps 1 (Environment Definition), 2 (System Requirements), 3
(Hazard Analysis), 4 (Safety Requirements), and wider organisational
and institutional failure.  The fatal 737 MCAS crashes are a typical
(but egregious) example: the likelihood and hazards of a single AoA
sensor fault were not dealt with.

It is extremely difficult to write and validate good safety
requirements and even harder to do hazard analysis.  Modern notations
\cite{Bhatt:FM22} allow some automated checks, but the overall process
requires human skill and experience.  Writing safety requirements
as constraints in an executable prototype is the wrong way to go as it
introduces premature concerns about implementation topics.

It is often possible to simplify matters by defining \emph{viability
domains} (regions of operation that maintain system safety) and
enforcing them and other safety properties using additional protection
systems or guards \cite{Deffuant11:viability}.  Examples of viability
domains are protection envelopes in nuclear plant that define
constraints on pressure, temperature, flux and flows (as opposed to
detailed sensing of the core), and in automotive systems they are a
combination of Operational Design Domains (ODDs) \cite[Section
6]{SAE-J3016} and the Safety Of The Intended Functionality (SOTIF)
\cite{ISO-PAS-21448,Schnellbach-etal19:SOTIF}.  Similar concepts are
used in economics (e.g., trading) and may be compared and contrasted
with nuclear safety protection \cite{Bloomfield&Wetherilt12}.  We
develop these ideas in Section \ref{syseng}, where we also
discuss adoption of the ``dependability'' perspective and ``defence in
depth.''

We then relate these engineering processes to assurance for AI systems in
general and Foundation Models in particular.  In outline, the problem
for AI systems in general is that they are built on machine learning or
neurosymbolic methods and we do not have strong knowledge of their
operation in any particular instance and so Steps 6 and 7 are difficult.
An additional problem for Foundation Models is that they are intended
as components in a wide range of applications and so it is difficult to
perform Steps 1 to 5.  Rather, these must be performed by application
developers, and we should ask what general claims about Foundation
Models will be of most value to them.  The next section considers these
topics in more detail.

\section{Preliminary Analysis}
\label{analysis}

We provide an analysis of whether there are concepts and techniques
from the critical systems perspective that might support the
development and implementation of Safety Frameworks.  We will present
the analysis under subsections with the broad headings: system
engineering, safety and risk analysis, implementation, and decision
analysis and support.  First, however, we consider aspects of the
report and commitments of the Seoul Conference from the systems
engineering viewpoint.

The interim report from the Seoul conference \cite{Seoul:interim23} is
comprehensive and identifies a broad range of risks.  It also,
correctly, points out that safety is a system property: the safety of
a mechanism such as AI must be considered in the context of the
environment in which it is to be deployed, as its risks (or, more
accurately, its hazards) are located in the environment.
Unfortunately, the report explicitly chooses not to address ``Narrow
AI'' that is ``used in a vast range of products and
services\ldots{}and can pose significant risks in many of them'' and
focuses on Frontier (i.e., advanced and ``wide'') AI.  It states that
this is due to ``the limited timeframe for writing this interim
report.'' However, we suggest that the capabilities of Frontier AI
have now reached a stage where they (or systems based on similar
technology) may be deployed in ``narrow'' applications in preference
to custom solutions, and hence the safety risks in these applications
should be considered part of Frontier AI safety.  This has two
implications: one is that the term ``safety'' should be interpreted in
its traditional sense as referring to any unintended consequence that
harms the system's environment; the other is that we must consider
consequences initiated by faults within the system (e.g.,
``hallucinations''), as well as those initiated in the environment
(e.g., by rogue users).  Thus, we find that the Seoul Report considers
an insufficiently broad range of systems and associated hazards.

Even within the coverage of the Seoul Report, the ``Seoul
Commitments'' focus narrowly on ``existential risks'' initiated by
malign users with access to hypothesized future systems with near-AGI
capabilities.  Again, we suggest the risk analysis is insufficiently
broad, even within the category of existential risks.  First, there
are significant societal risks from near-term capabilities that we
term AFGI: Artificial Fairly General (or Fairly Good) Intelligence.
These include unemployment due to systems that are just ``good
enough,'' degraded job performance due to them actually not being very
good, proliferation of mediocre information leading to ``knowingness''
\cite{Malesic:knowingness23,Lear:open-minded99} and mistrust of
institutions, and so on.  These have the potential to reverse decades
of human progress, yet none of them require actively malicious users:
imperfect AI will do it.

We will discuss the well-established traditional processes of risk
assessment and systems engineering below, and then attempt to relate
them to Frontier AI and the Seoul Commitments.   We will also discuss
the traditional methods of risk mitigation.  These require
identification of hazards and formulation of explicit requirements
concerning their mitigation.  Like hazards, safety requirements
concern (the system's impact on) the environment.  The mechanisms by
which the system will achieve its requirements are described in its
specifications and these must be shown to ensure the requirements; the
system implementation must likewise be shown to ensure its
specifications.  The latter demonstration is largely based on an
understanding of how the implementation works in every circumstance,
which is seldom feasible for learned behaviours.

Requirements and specifications concern the whole system, although they
may be decomposed into sub-specifications for its components.  If only
weak assurance can be provided for a component, then the system
architecture must generally provide some other component that
compensates for this: for example, a ``guard'' that performs strongly
assured runtime checking.  Many such architectures are possible but
``diversity'' and ``defence in depth'' are very common architectural
strategies that combine as the ``Swiss Cheese Model,'' which is
discussed by the Seoul Report \cite[Section 5.1.2]{Seoul:interim23}.

None of the corporate or national frameworks that we have examined
make any mention of these topics.  Instead, only weak forms of AI
assurance are considered (e.g., ``red teaming''), and mitigation
consists in adjustment to the AI model (e.g., ``fine tuning''), which
resembles the fox guarding the henhouse.

An exception is the proposal for ``Guaranteed Safe'' AI
\cite{Dalrymple-etal24:GS-AI} and the associated ``Safeguarded
AI'' program from the UK ARIA
(see \url{https://www.aria.org.uk/programme-safeguarded-ai/}):
\begin{quote}
``We introduce and define a family of approaches to AI safety,
collectively referred to as guaranteed safe (GS) AI. These approaches
aim to provide high-assurance quantitative guarantees about the safety
of an AI system's behaviour through the use of three core
components---a formal \emph{safety specification}, a \emph{world
model}, and a \emph{verifier}. We will argue that this strategy is
both promising and underexplored, and contrast it with other ongoing
efforts in AI safety.''
\end{quote}
But even here we see only a partial recognition of the full process of
critical systems development: it addresses only three of the eight
steps described in the previous section (steps 1, 4, and 7).  It is
possible the other steps are implicit but we argue that they are
broken out in our summary for good reason: in particular hazard
analysis, requirements validation, and the overall assurance case
(Steps 3, 5, and 8) are the most important, difficult, and
failure-prone of all the processes and must be given explicit
attention.

We accept that it is difficult to apply the traditional approach to
Foundation Models and existential threats: typically, the environment
in which the Foundation Model will be deployed is unknown, as is the
surrounding architecture and selection of potentially mitigating
components.  Furthermore, at present, we lack detailed understanding
of how a learned model works, and of its properties.  Nonetheless, as
with risk assessment, we will attempt to relate traditional
understanding of risk mitigation and assurance to Frontier AI and the
Seoul Commitments.

\memo{Below, we surely address more than just steps 1 and 2; I think
this paragraph needs rewriting to add hazard analysis, definition and
validation of safety requirements, and decision analysis and
communication}

Our approach to supporting AI Safety Frameworks is to reframe the
problems to be analysed by considering what is the system, how broad
do we draw boundaries, how do we decide on what are tolerable risks,
and by whom and under what circumstances are they tolerable.  These
considerations need to address not just a safety focus on the loss
event, no matter how catastrophic it might be, but also a resilience
focus on potential recovery before and after the situation escalates.

\subsection{System Engineering: What is the System?}
\label{syseng}

The first point we wish to emphasize, and it is recognised in the
Seoul Report, is that safety and similar concerns are system
properties.  A system has a mechanism and an environment or context.
In general usage, the mechanism is often referred to as ``the system''
and we will do the same, but it must be understood that discussion of
safety only makes sense in conjunction with an environment.  This is
because discussion of safety requires consideration of the hazards
that the system may pose, and these are all in the environment.  It is
for this reason that the only things certified by the FAA are
airplanes and engines (and propellers): software is not certified
separately but only in its role as part of a specific function in a
specific airplane intended to operate in a specific context.  Of
course, software is expected to possess some quality attributes as
part of its general development quite apart from those specific to its
context of use, just as the metal used in some component is expected
to be of high quality.  For software, these include good development
practices, configuration management, disciplined coding practices,
static analysis, testing, and so on.  We find that much of the
discussion of assurance for AI software is of this general kind, and
lacks full consideration of the specific system context, hazard
analysis, requirements validation and so on.

\subsubsection[Boundaries, socio-tech issues, and overall service perspective]{Boundaries, socio-technical issues, and overall service perspective}
\label{boundaries-socio-technical-issues-and-overall-service-perspective}

In defining the safety properties and associated hazards we need to
explore system boundaries.  A cause of failure in complex systems is to
draw the system boundary too narrowly and consider just ``equipment'' or
in AI the ``algorithm'' rather than considering the overall
socio-technical system that is important for delivering the service
 \cite{Baxter&Sommerville:socio13}.

People are part of the wider system and may be a threat due to
malicious behaviour or have unintentional impacts through human
behaviour and error.  The role of the individual in achieving safety
is addressed by human factors in a variety of industries
\cite{ONR-HF-TAG} but this guidance typically focuses on the trained
operator---where we also need to consider use by the general public
and disparate sets of users.  Broadening out from human factors we
need to consider how people adapt to the system, and use it in
off-label or imaginative ways.  In terms of people and organisations
there are two complementary views: Normal Accidents \cite{Perrow84}
and High Reliability Organisations \cite{Rochlin93} that have both
been explored in the context of nuclear weapons safety \cite{Sagan93}.

In terms of AI as a service, we must also consider the delivery of
this service and the possible harms and failures from running a
globalised infrastructure (e.g., how are updates distributed---recall
the CrowdStrike crash).  Evaluation of these might seem more mundane
than new AI capabilities but they must be assessed as a source of
risks and will also consume a risk budget, thereby increasing the
criticality assigned to other (AI) aspects.

\subsubsection{Architecture: use of models, guards, and defence in depth}
\label{architecture}

To design and assure systems we need validated models (in the sense of
abstracted descriptions of how things work).  The models need to be
valid for the task they are being used for.  At the very least, we
require adequate models for safety analysis, and these can be implicit
or explicit or both (e.g.  an explicit system architecture diagram
with implicit behaviours that are assessed by experts in a hazard
analysis activity).  To simplify the modelling task and make it
feasible, we may constrain the world to reflect our models (e.g., make
it synchronous or time triggered \cite{Kopetz&Steiner22}), or design
systems so they can be modelled.  The role of models in engineering is
neatly summed up by paraphrasing \cite{Lee17:Plato}: scientists use
models to understand the world, engineers to change it.

To simplify modeling we can define viability domains: regions of
operation for the system where safety is maintained.  These allow us
to model safety of the system with limited understanding of its
components and also reduce the sensing challenges as we only need to
sense approximate and external values.  For example, an animal's
health may be monitored by body temperature, respiration rate and
heart rate, and the state of a nuclear reactor core by only a few
measurements of flows, temperature, nuclear activity and pressures: we
do not need detailed modeling and intrusive monitoring of internal
details.  Viability domains and safety properties can then be enforced
using relatively simple external protection systems known as guards or
monitors to provide ``runtime verification''
\cite{Falcone-etal13:runtime-verif,Rushby:RV08}.
The ``dependability'' perspective uses suitable architectures to
further reduce the criticality of guards' safety properties and can
apply these ideas recursively, providing defence in depth over
multiple layers (e.g., for the guards themselves)
\cite{Bloomfield&Rushby:Assure24,Bloomfield&Rushby:AI24}.

\subsection{Safety and Risk Analysis: How Good Does it Have to Be?}
\label{safety-and-risk-analysis-how-good-does-it-have-to-be}

We acknowledge that for generic AI components, and foundation models
in particular, the eventual applications and their environments may be
unknown and so it is hard to perform Steps 1 to 5 of the engineering
and assurance outline presented earlier.  A plausible approach is to
consider worst-case possibilities, and the Seoul Commitments appear to
do this: namely, hostile environments, highly hazardous applications,
and catastrophic consequences.  However, it is not articulated how
these specific applications, hazards, and environments were selected.
Other well-cited studies consider different applications such as
healthcare, law, and education, and identify very different hazards
such as threats to fairness, environment, and economics
\cite{Bommasani21:foundation-risks}.

AI Assurance could fail in some of its purposes if the risks addressed
do not coincide with societal concerns.  Hence, we suggest a
systematic and open assessment should be undertaken of hazards and
associated risks in a wide range of potential applications, and the
judgments ``catastrophic'' and ``existential'' should be carefully
delineated.  And we suggest that corporate and national commitments
should focus on representative risks and not only those considered
worst-case.  In particular, the impact of everyday AI as a
force-multiplier should be kept in mind: minor risks may become
intolerable when replicated on a vast scale.  Furthermore, those
representative cases should be examined in the framework of the
traditional 8-step safety engineering process, particularly the
performance of hazard analysis (Step 3) and the construction (Step 4)
and validation (Step 5) of safety requirements.  It is only by
considering representative cases in some detail that we can identify
whether proposed assurance techniques are likely to be adequate or
beneficial.

Observe that concerns in this section stem from the genericity of
foundation models and their lack of a defined context, while concerns
in Section \ref{implementation} on implementation stem from its basis
in machine learning.

\subsubsection{Design basis events and threats}
\label{design-basis-events-and-threats}

In defining safety requirements and associated hazards we need to
address a multiplicity of different environments and events.  One
approach to exploring this issue is to consider a range of scenarios,
as has been done for the undermining of democracy and for Chemical,
Biological, Radiological, and Nuclear (CBRN) applications.  But if we
are to move from illustrative scenarios to a necessary and sufficient
a set of events to be addressed, we need to model this infinite set of
possibilities.  One approach, common to evaluations in engineering
complex systems, is to define justified worst-case events that bound
the space (e.g., the biggest projectile crash on the reactor building,
or the largest credible tsunami) or to have distributions of these
events.  These events are known as the Design Basis Events
\cite{ONR-Generic-Assessment}.  Of course, there are many epistemic
issues here in judging what are credible events, and characterizing
our uncertainty in the world model they reflect, but they do provide a
systematic way of addressing the multiplicity of environments and
events.  Events can occur along many dimensions (e.g., projectile
impact, earthquake, flooding) and for AI applications it may be
difficult to enumerate a set with adequate coverage.

\subsubsection{Integrated safety and security}
\label{integrated-safety-and-security}

Similarly, when we talk about security, we need an environment context
that describes the threat actors and their interaction and potential
harm to the system.  Design Basis Events can be generalised to or for
security as Design Basis Threats \cite{ONR-Generic-Assessment}.  There
is an interesting symmetry where safety can be seen as the possible
impact of the system on its environment and security as the impact of
the environment on the system.  Security attacks can also lead to harm
to the environment (whether in terms of classical safety, or release
or compromise of information).  There is work on the integration of
functional safety and security with guidance published by the UK
National Protective Security Authority (NPSA)\footnote{See collection
of material on security-informed safety at
\url{https://www.npsa.gov.uk/security-informed-safety}.} with the
slogan ``if it's not secure it's not safe.''

\subsubsection{Identification and shaping of risk and tolerability}
\label{identification-and-shaping-of-risk-and-tolerability}

Safety and risk analysis derive the safety properties and functions
required of the system but we also need to consider how good does the
system have to be, and how confident do we need to be about this.  The
first of these is often captured in terms of criticality levels and the
probability of something bad happening: for a plane crash the ``bad''
can be obvious, a major loss of life, but its tolerable likelihood is a
more complex socio-technical question.  This can be expressed both
qualitatively and quantitatively (e.g.  in terms of probability of
failure on demand, or accident frequency), depending on the safety
property.

Deciding criticality or safety levels, or tolerability of risk, is a
social and political issue and involves an analysis of who has these
risks and who has the benefits.  In the UK these were brought together
by a public inquiry into nuclear safety and resulted in the UK safety
agency, the HSE, whose publication on ``Reducing Risks, Protecting
People'' is a document which describes its decision-making process
(known as r2p2 \cite{HSE:r2p2}).  Societies' risk values vary across
technologies and societies: being killed by cars seems more tolerable
than by aviation or rail; the US tolerates gun deaths while many other
societies do not.  The tolerability of being killed by autonomous
vehicles has been judged to be 100 times less than for accidents
initiated by a human driver.  Suffice to say that the risk appetite
for harm from AI can be shaped by government policy but it is likely
to be dynamic, and will vary across technologies, applications, and
societies.

To address tolerability, we need to be able to consider the evolution of
risk over a very uncertain future and consider the aggregation of risks
and to whom they apply.  There are value judgments to be shaped and
elicited on topics such as how a large volume of small harms (e.g.
distress caused by widespread interaction with an LLM) can be compared
with risk from, say, an AI-controlled chemical plant causing pollution
and injury.  To address this, the analyses arising from AI safety
frameworks must make the risks clear and communicate them accordingly
(see role of cases in Section \ref{decision}).

\subsubsection{Recovery, resilience and adaptation}
\label{recovery-resilience-and-adaptation}

In national critical infrastructure risk assessment, it has been
recommended \cite{RAE:resilience} that we recognise both chronic and
acute risks and their interplay.  Chronic risks, such as a lack of
social cohesion, undermining trust in institutions, and long term
cognitive changes, are all significant factors in societal risk and
social acceptability of AI\@.  Another lesson from national critical
infrastructure assessment is the need to consider resilience and
recovery: by explicitly considering resilience, some harms may be
recoverable with tolerable losses while others may lead to long term
toxicity, or once occurred cannot be recovered (e.g., loss of secret
information).

Resilience is most broadly defined as the capacity of a system to
return to its original state after shocks.  It can be useful to
distinguish two subtypes within this \cite{Bloomfield&Gashi08}.

\begin{description}
\item[Type 1:] resilience to design basis threats and events.  This
could be expressed in the usual terms of fault-tolerance,
availability, robustness, etc.

\item[Type 2:] resilience beyond design basis threats, events and
use. This might be split into known threats that are considered
incredible or ignored for some reason, and other ``black swan''
threats that are true unknowns.
\end{description}

Often we are able to engineer systems successfully to cope with Type 1
resilience using methods of redundancy and fault tolerance.  Type 2
resilience is a more formidable challenge.  We may choose to make
systems more heterogeneous and interconnected and with more resources
to support the second type, but doing so might make them more
expensive and suboptimal in terms of the first type of resilience.

\begin{figure}[t]
\centering{\includegraphics[width=4.5in]{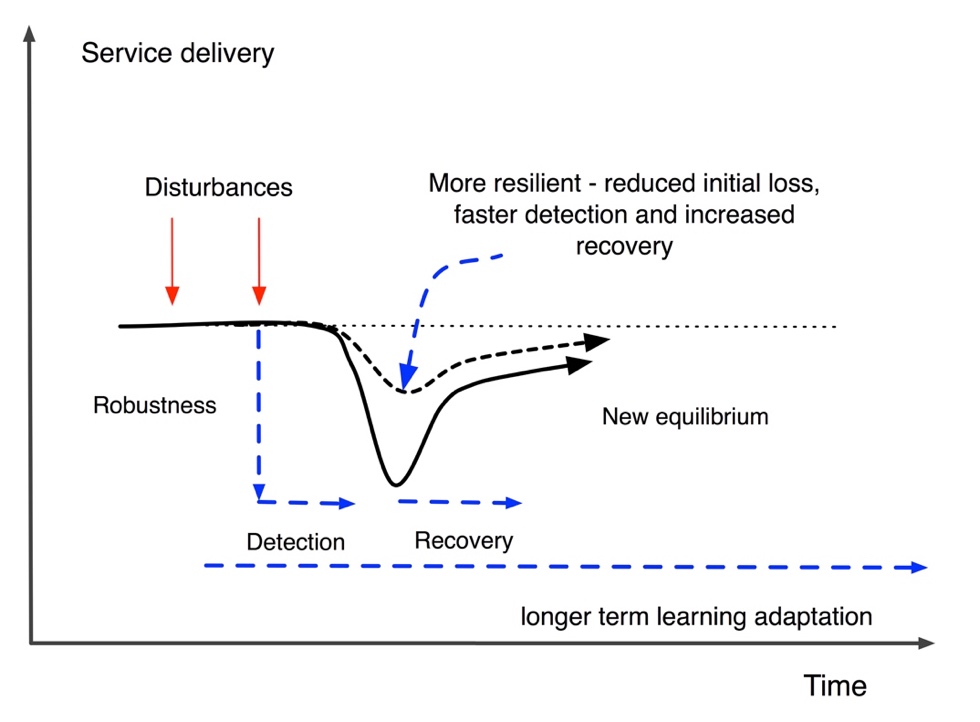}}
\caption{\label{resilience}Resilience}
\end{figure}

Furthermore, complex systems can be challenged not just by exogenous
events but also by those internal to the system: some of these might
be the traditional failure of components (either technical or human).
However, increasingly, it is claimed, significant failures are due to
an accumulation of normal variability that in some instances become
correlated and, because of the very non-linear response of the system,
leads to unexpected and/or unwanted behaviours: so-called
\emph{emergent misbehaviours} \cite{Mogul06:emergent-misbehavior}.  In
this last scenario, reductionism is a much less successful strategy,
and a more holistic approach is necessary or even essential.

The ``normal accidents'' school identifies \emph{tight coupling} and
\emph{interactive complexity} as key organizational factors in
accidents that runaway into catastrophes \cite{Perrow84} and these
factors extend to computer systems \cite{Rushby94:Taxonomy}.  It is
crucial to ensure that mechanisms for resilience serve to break tight
coupling and interactive complexity, not add to them.

The longer time horizon must also be considered.  It is clear AI can
have a major impact on society, on how we engineer complex systems,
and on the nature, and type of risks entailed.  As AI develops it will
be both a source of risk but also a key resource to understand and
mitigate risk.  It is not clear what the risk profile will look like
over time but the classic questions of who benefits and who loses are
relevant, along with issues of generational fairness.  There are
likely to be many different voices and value sets so that the use of
assurance cases to communicate and reason will be important as will
the use of explicit counter-cases and other dissenting cases (see
Section \ref{decision}).

One framework that can be used to describe how systems evolve and
adapt is the ``Open Systems Dependability'' perspective that
originated in Japan from their dependability research and approach to
consensus management \cite{DEOS13}.  This might impact, for example,
the safety case approach by causing it to consider explicitly how
robust the case is to changes, how it might detect changes, and how
the assurance argument and system will adapt to new circumstances.

\subsubsection{Summarising and communicating the dependability strategy}
\label{summarising-and-communicating-the-dependability-strategy}

\begin{figure}[ht]
\centering{\includegraphics[width=4.65812in,height=2.34797in]{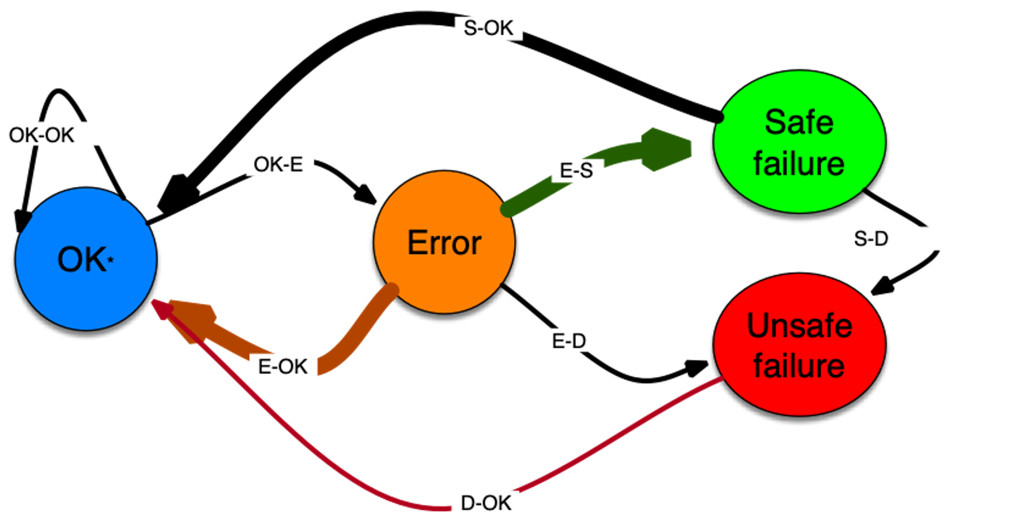}}
\caption{\label{four-state}Four State Model for Dependability}
\end{figure}

In engineering critical systems, we often use a 4-state model (see
Figure \ref{four-state}) to describe the different approaches to
achieving dependability.  To achieve dependable systems, we can
minimise the transition from the OK to error state and we can have
fault tolerance and management that return the system to an OK state
without loss of service.  If the error state escalates, we can design
the system so that it fails to a safe or minimum loss state and then
recovers.  If it does fail and leads to loss, then we can plan and
perform incident recovery.  The balance between these transitions
varies from system to system.  This model can be applied recursively
to components within the system design.

For critical systems, AI needs to focus not just on keeping operation
within the OK state but should consider the other transitions in this
dependability model.  Design that takes recovery into account is key to
achieving resilience.  We suggest this may be best performed in an
architecture with guards and defence in depth.

\subsection{System Implementation: Impact of Criticality on Development}
\label{implementation}

Having considered what exactly is the system and how much we should
trust it, we now consider its implementation.

For a system implementation to be considered safe, we need to perform
system safety engineering so that it \emph{is} safe for the identified
hazards, and we need to provide assurance that this is so.  These
correspond to Steps 1--6 and the accompanying parts of Step 8 in the
engineering and assurance outline presented earlier.

Safety engineering attempts to eliminate hazards (e.g., if fire is a
hazard, then remove flammable material and sources of ignition), or to
mitigate them (e.g., add a fire extinguishing system, but then failure
of that system becomes an additional hazard).  Systems engineering
results in the specification (Step 6), which is then realized by the
implementation (Step 7).  Assurance for the implementation is generally
based on strong understanding of how it works and how it interacts with
its environment, as described in its specification.  When the system (or
component) implementation is software, assurance typically aims to
show that it is \emph{correct} with respect to the specification.  If
definitive proof of correctness (e.g., formal verification) is
infeasible, then several weaker methods (e.g., testing, static analysis,
manual review) are used in combination.  The relevant parts of the
overall assurance case will document these methods and any caveats or
concerns.

The overall assurance case should deliver two judgments: how safe is
the system, and how confident can we be in our assessment of it.  The
first is reliability wrt.\  critical failures and is generally
represented as a qualitative or numerical reliability measure (e.g.,
``not expected to occur in the lifetime of all aircraft of one type,''
or ``probability of failure less than $10^{-9}$ per hour, sustained
for 20 hours'').  The second can be represented as a subjective
probability (e.g., 95\% confident, which can be interpreted as an
expectation of being wrong one time in twenty).  The first is
evaluated by the methods of reliability analysis, the second by expert
judgment and associated techniques.

When either of these measures is incommensurate with the severity of the
hazard (i.e., its risk), then the system engineering may need to be
revised to mitigate the concern.  In the case of software, this may
require a more complex architecture.  For example, we may provide a
highly assured \emph{guard} (or \emph{monitor}) that performs runtime
checking on the behaviour of the primary software component.  This is
appropriate when it is difficult to construct a result, but checking it
is simpler.  This architecture can increase both the critical reliability
(i.e., safety) of the system and confidence in its assurance.  In other
circumstances, we may replicate the primary system component with a
\emph{diverse} secondary (or several secondaries) and compare results
(in AI this is often known as an ensemble or portfolio architecture).
``Diverse'' means the secondary should be developed independently of the
primary.  There is little doubt that diversity increases reliability
overall, but it is very difficult to quantify by how much, or to provide
assurance that it delivers useful benefit in any particular instance
(you cannot simply assume that failures are independent).  Hence, this
architecture probably increases safety significantly, but not so much
confidence in assurance.

In the case of components that use machine learning, and Foundation
Models in particular, it is currently difficult to achieve any detailed
understanding how the system works.  We are therefore forced to use black
box methods of assurance such as random testing, red-teaming, robustness
checks and so on, which deliver very low confidence.  It is therefore
necessary to consider architectural methods for assurance such as highly
assured guards or diverse secondaries.  Due to the wide potential
application of foundation models, implementation assurance has to be
performed in the absence of a system specification and for this reason
we advocate consideration of representative applications, as discussed
in Section~3.2 so that recommendations and commitments can build on
experience.

As noted above, how confident we need to be in a safety property will
typically vary by orders of magnitude between everyday systems and those
that might pose significant harm.  This has a profound impact on the
engineering of critical systems: to move from everyday reliability for a
low harm system to high reliability for a critical one needs a different
approach to engineering.  It is not sufficient to just try harder or
select the upper part of the distribution of development approaches:
critical systems require different methods of development and
justification.

There are minimal requirements for us to have trust in any system: basic
quality procedures and configuration management so that we know what the
system is, what the evidence relates to, and whether we have a
consistent set of development artefacts that we know the provenance of.

Having considered what exactly is the system, how good it has to be, and
the impact on implementation, we now consider how much we should trust
it.

\subsection{Decision Analysis and Support: How Much Should we Trust?}
\label{decision}

Assurance serves (at least) three purposes: it helps the developers
ensure that their system is safe, it provides a basis whereby external
evaluators can assess and approve deployment of the system, and for
those who do deploy the system it communicates reasons for trust, the
assumptions and limitations of that trust and, hence, conditions on the
context of deployment.  Ideally, these activities proceed in parallel as
parts of the co-design of systems and their evaluation.

Given how broad the application of AI can be, there is a need to
define carefully what decision an assurance case is supporting and to
understand the needs of the stakeholders in communicating the story
that the case is telling.  Cases can support both risk communication,
and the building of trust and elicitation of values.  As we have
discussed above, there can be a wide range of views (see debate on
existential risks) and so counter-cases (those that argue a contrary
or negative claim) can have a role in understanding and explaining the
different perspectives.  There is potential benefit in a case (or set
of linked cases) that addresses the socio-technical aspects,
integrates impact of security on safety, and addresses resilience and
adaptation.  Taken together these factors can calibrate the rigour
needed in the case.

\subsubsection{Assurance cases for reasoning and communication}
\label{communication}

As already mentioned, assurance cases (generalizing the earlier notion
of a safety case) provide a framework for constructing---and a lens for
viewing---assurance, and are recognized as a potential approach within
the AI community.  However, we find their application is often less
sceptical than we would prefer.  The primary hazards in assurance are
complacency and confirmation bias.  Those who construct, review, and use
assurance cases would do well to recall Lakatos' dictum: ``The virtue of
a logical proof is not that it compels belief but that it suggests
doubts'' \cite[page 48]{Lakatos}.

From our experience with the demands for innovation in complex
safety-critical systems, we have been developing an approach dubbed
``Assurance 2.0'' that supports sceptical analysis and is being
transitioned in a number of application areas (see collection of
papers on Assurance 2.0 \cite{Assurance2-home} and
\cite{Bloomfield&Rushby24:CBJ} in particular for references to the
technical and scientific terms used in the following paragraphs).

Assurance 2.0 provides a framework for assurance around \emph{claims},
(structured) \emph{argument}, and \emph{evidence} (building on the
existing CAE approach).  Arguments are constructed from just five
building blocks or steps (concretion, substitution, decomposition,
calculation, and evidence incorporation), which reduces the
``bewilderment of choice'' in free-form arguments.  Argument steps are
generally expected to be deductive: that is to say, the conjunction of
child subclaims to each argument step must entail the parent
claim---because otherwise there is a ``gap'' in the reasoning.
Side-claims (logically no different to other subclaims, but conceptually
distinct) factor out deductiveness conditions (e.g., the subclaims
partition the parent claim, or the parent claim distributes over
components enumerated in the subclaims).  Furthermore, we set a high bar
for accepting argument steps and the overall conclusion: they must be
\emph{indefeasible} (a notion from epistemology), meaning that we cannot
conceive of any new information that would change our judgments.

The sceptical quest for deductiveness and indefeasibility is supported
by \emph{defeaters}, which are claims that express a doubt and can
target any node in the argument.  Defeaters can be sustained or refuted
by a subargument just like other claims, and can be retained (but
typically hidden and optionally revealed) in the case as they can assist
later developments, and can also help evaluators who may find that their
own doubts have already been considered.  All defeaters must be refuted
(or accepted as \emph{residual risks}) in a finished case.  Defeaters
also support an alternative approach called \emph{eliminative
argumentation} where, instead of confirming a positive claim (e.g., the
system is safe), we refute a negative one (e.g., the system is unsafe).

Evidence is also evaluated sceptically, using the ideas and measures of
\emph{confirmation theory} (which come from Bayesian Epistemology).  That
is, we do not merely ask whether evidence supports a claim, but how much
it adds to our prior belief in the claim, and whether it also supports
alternative claims and the counterclaim in particular.  We generally also
distinguish between the \emph{measured} claim supported by evidence
(e.g., ``we performed MC/DC testing and discovered no errors'') and the
\emph{useful} claim derived from it (e.g., ``we have no unreachable
code''); confirmation theory is applied to the useful claim.  The step
from measured to useful claims is performed by a substitution block that
typically cites some \emph{theory} (e.g., the theory of MC/DC testing).

Theories are descriptions of some standard subargument, preferably
culminating in a parameterized (and ideally, pre-approved) subcase that
can serve as a template to be instantiated or referenced in some larger
assurance case.  Selection and instantiation of theories can be partially
automated, and the Clarissa toolset that supports Assurance 2.0 has a
\emph{synthesis assistant} for this purpose.  Both synthesis and manual
construction of arguments are assisted when claims use standardized
terminology and are phrased in a consistent style.
\emph{Autoformalization} using LLMs can be a great help here.

Those who deploy a system do not merely want to know that it is declared
to be safe: they want some idea why it is safe, why they should have
confidence that this is so, in what environment they should deploy it,
and how they should use it to ensure safety.  An assurance case can
provide this information and should be made available to end users, but
assurance cases are often very large so we generally expect evaluators
to provide a \emph{sentencing statement} that provides this information
in succinct form.  Many assurance cases can be summarized by enumerating
the theories that they use and the overall structure of their use.

As noted above, current AI evaluations use a combination of random
testing, red-teaming, robustness checks and so on, which deliver very
low confidence.  There is a need for theories and new analysis
techniques that allow an extrapolation from coverage and reduced
models to deployed ones.  In Assurance 2.0 terms, we need theories
that allow us to go ``from something measured to something useful.''

\subsubsection{Identify decision criticality not just system criticality}
\label{identify-decision-criticality-not-just-system-criticality}

We need to consider how good the evaluation has to be, and this is of
course linked to the decision being made.  We need to understand that
decision and, as with systems, consider the resilience aspects: how
much harm might done before we can detect a bad decision and whether
we can recover from such a bad decision.

As noted above, the tolerable failure rates for safety properties will
vary by orders of magnitude between everyday systems and critical
systems that might pose significant harm.  Similarly, the confidence
required in our evaluation will also increase as the criticality of
the system increases.  A qualitative illustration of this is the
difference between a cut down CAE approach to confidence building for
resilience of commodity devices (such as home fridges) using
Principles Based Assurance (PBA)\footnote{UK National Cyber Security
Centre
\url{https://www.ncsc.gov.uk/blog-post/making-principles-based-assurance-a-reality}.}
contrasted with the rigors of an Assurance 2.0 case for a safety
critical system.  Some illustrative quantitative modelling of how
confidence increases with criticality is provided in
\cite{Bishop-etal:TSE11}.

\subsubsection{Explicit approach to confidence in safety claims}
\label{explicit-approach-to-confidence-in-safety-claims}

There are a number of technical approaches to evaluating how confident
we are in safety claim: one is to use a structured approach to
modelling the justification \cite{Bloomfield&Rushby24:CBJ} in which
confidence in parts of the evaluation can be combined in a
conservative manner.  This could be by approximate worst case
propagation of doubts or by the use of more nuanced theories that
explicitly deal with confidence.  There are also technical methods
like the ``chain of confidence'' that help model the impact of being
wrong by modelling how wrong we might be, and applying this
recursively (see \cite{Bloomfield&Rushby24:CBJ} and
\cite{IAEA18:nuclear}).

\subsubsection{Explicit approach to judgment bias}
\label{explicit-approach-to-judgment-bias}

Assurance 2.0 provides an explicit approach to addressing confirmation
bias through the use of defeaters, confirmation theory, and explicit
counter cases.  This can be augmented with surrounding processes that
also provide for independence and diversity of opinion.

A different kind of bias arises when we fail to consider alternative
cases that might be as technically sound, but based on different values
and judgements.  The techniques we use to address bias can also
contribute to understanding these different approaches and could be part
of building consensus as well as respect for different positions.
We therefore envisage a range of cases that are used to articulate and
communicate different viewpoints.

\subsubsection{Distinguish different types of argument and inherent strengths}
\label{distinguish-different-types-of-argument-and-inherent-strengths}

A high-level factoring of argument approaches is to use the ``strategy
triangle'' that describes justifications in terms of rule-based,
goal-based, and risk-informed approaches that focus on compliance,
behaviors, and vulnerabilities, respectively
\cite{IAEA18:sw-dependability}.  For systems that pose very
significant potential harm all three aspects will be relevant with
those addressing behaviours more compelling.  When dealing with
extreme behaviours (e.g.  catastrophic failure of a nuclear reactor
pressure vessel) arguments about the incredibility of failure may
combine deterministic, analytical and probabilistic reasoning.  In
general, analytic arguments are stronger than probabilistic ones.  In
Assurance 2.0 we have an approach that combines them where the
deductive part is supported by inductive argument about the
assumptions.  A simple example is the unfounded fear that a civil
nuclear reactor might cause a nuclear explosion.  The argument that by
design there is not enough fissile material in the core is much
stronger than a probabilistic argument in which the core has
significant fissile material, but we rely on probabilistic evaluations
to show it is very unlikely that a core meltdown will produce a
critical configuration.  Similar examples come from computer science:
a proof of absence of critical defects, with assumption doubts
addressed, is stronger in principle than statistical testing because
it covers all conditions.

\subsubsection{Automation and tempo}
\label{automation-and-tempo}

The abundance of uncertainties on the evolution of capabilities,
knowledge of risks and benefits, and attitudes to risk tolerability
all emphasise the need to frequently update individual assurance
cases.  The need for greater tempo in the use of cases has been
apparent for some time with the need for innovation to support the
``compile to combat'' doctrine and DevSecOps.  The DARPA ARCOS
programme sponsored a number of projects on automation of
certification.  We were part of the Clarissa project building on the
Adelard ASCE platform.  The program adopted a ``documents as data''
paradigm where legacy and new analysis methods updated a semantic web
triple store.  This was then used by assurance case tools to feed
evidence into case construction and analysis tools
\cite{Varadarajan-all:DASC24}.
Although some of the tooling is at low TRL, the lessons learned show
that greater automation is feasible, and this should be embraced.

\section{Evolution of AI Safety Frameworks}
\label{evolution}

It is recognised that the development and implementation of AI Safety
Frameworks would benefit from adopting and adapting best practices
from other industries.  A first step towards this is to identity where
there might be fruitful areas of interest that can be developed
further, building on success, failures and open issues within the
dependability engineering of critical systems.

We have provided an initial analysis driven by following questions
\begin{itemize}\itemsep=0pt
\item
What is the system?

\item
How good does it have to be?

\item
What is the impact of criticality on system development?

\item
How much should we trust it?

\end{itemize}

\begin{figure}[th]
\centering{\includegraphics[width=5.5in]{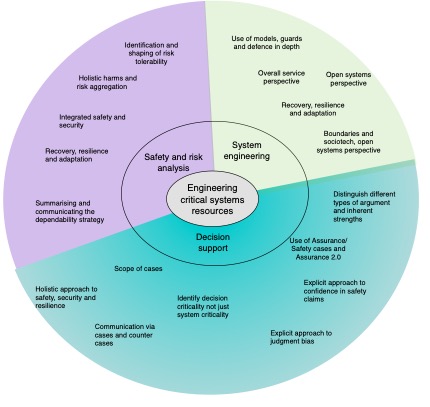}}
\caption{\label{ideas}Ideas from Engineering Critical Systems}
\end{figure}

The topics we have identified can be grouped according to whether they
address system engineering, risk analysis, or decision analysis and
support.  The topics are summarised in Figure \ref{ideas} and are as
follows.
\begin{description}
\item[System engineering]\mbox{}
\begin{itemize}\itemsep=0pt
\item
System evaluation and compositional assurance

\item
Boundaries and sociotech, open systems perspective
\item
Overall service perspective
\item
Use of models, guards and defence in depth
\item
Recovery, resilience and adaptation
\end{itemize}

\item[Risk analysis]\mbox{}

\begin{itemize}\itemsep=0pt
\item
Identification and shaping of risk tolerability

\item
Holistic harms and risk aggregation

\item
Design Basis Events and Threats

\item
Integrated safety and security

\item
Recovery, resilience and adaptation

\end{itemize}

\item[Decision analysis and support]\mbox{}

Scope and use of assurance cases
\begin{itemize}\itemsep=0pt
\item
Holistic approach to safety, security and resilience
\item
Communication via cases and counter cases
\item
Use of cases to understand disparate views
\end{itemize}

Use of Assurance/Safety cases and Assurance 2.0, in particular

\begin{itemize}\itemsep=0pt

\item
Claims, Arguments and evidence (CAE) and CAE Blocks

\item
Practical Indefeasibility and deductive core

\item
Explicit approach to judgment bias

\item
Explicit approach to confidence in safety claims

\item
Identify decision criticality not just system criticality

\item
Distinguish different types of argument and inherent strengths

\item
Use of automation and increased tempo

\end{itemize}

\end{description}

We suggest that AI safety frameworks should consider and address all of
these topics in more detail than at present.

There are a number of important ways in which we think that AI Safety
Frameworks need to change:

\subsubsection*{The importance of resilience of the system and the
decision making}

There is a need to consider resilience and recovery: by explicitly
considering resilience, some harms may be found to be recoverable with
tolerable losses while others lead to long term damage, or once
occurred cannot be recovered.  We need to understand the criticality
of decisions (e.g., to deploy) and, as with systems, consider the
resilience aspects: how much harm is done before we can detect a bad
decision and whether we can recover from it.

\subsubsection*{Hazard and risk analysis and risk tolerability}

As AI systems become capable of posing potentially intolerable levels
of risk there will be ever more need for rigorous hazard analysis to
determine the true extent of present and future risks, and whether
mitigations should be sought in the AI mechanisms themselves, or in
the larger socio-technical system.  We are concerned that if the
system boundaries are not broad enough, the nature and tolerability of
the risks will not be sufficiently elaborated.

\subsubsection*{Development and evaluation methods for more critical
systems are vastly different from lower criticality ones}

Development methods do not scale from everyday systems to critical
ones.  Safety with regard to extreme risks requires failure rates
(e.g., $10^{-9}$ per hour) that are outside individual human
experience and can only be achieved by the most disciplined (and
typically conservative) systems engineering and massive investment in
assurance.  There are likewise issues in scaling the confidence in
evaluation that is needed to support the decision to operate.
AI evaluation methods lack theories that allow behaviours to be assured
with quantifiable confidence.  We are concerned that the challenges of
truly critical AI systems may be underestimated.

There will therefore be a need for enhanced rigour in systems, risk,
and decision engineering as AI is employed in increasingly significant
applications.  Furthermore, unwelcome risks may be posed by unwise
deployment of fairly good near-term AI and these need to be considered
alongside existential risks of hypothetical AGI.

\subsubsection*{Systems perspective}

Because risks arise from the system context, AI safety frameworks should
adopt a system perspective and be developed in collaboration with
application developers to identify architectures in which the AI is
buttressed by diverse replicas, guards, and other mechanisms for
mitigation and confidence building.  Confirmation bias must be avoided
and assurance should adopt a rigorous and highly sceptical perspective.
In particular the following topics should be considered.

\begin{itemize}\itemsep=0pt
\item
  Use of system models (in the sense of formal descriptions), rigorous
  implementation of guards, and defence in depth,

\item
  Reasoning and autoformalisation in decision support,

\item Increased use of challenge and counter-cases, and explicit
  seeking of contrary evidence.
\end{itemize}

\section{Summary and Conclusions}
\label{summary-and-conclusions}

There are three main ways in which AI Assurance might go wrong: one is
that it addresses the wrong risks, second is that its techniques are
inadequate for those risks that are addressed, and finally it may fail
to communicate its claims effectively to the public and other
stakeholders.

We summarised how safety engineering is performed in traditional
critical systems, such as aircraft flight control.  We provided an
analysis of whether there are concepts and techniques from this
critical systems perspective that might support the development and
implementation of AI Safety Frameworks.  We presented the analysis
under the broad headings: system engineering, safety and risk
analysis, implementation, and decision analysis and support.  We have
been selective in this analysis, making a judgment of what we think
might be fruitful at this stage in the development and implementation
of AI safety frameworks.

Our analysis considered four key questions: What is the system? How
good does it have to be? What is the impact of criticality on system
development? How much should we trust it?  We identified a number of
topics we think worthy of further discussion.  In particular, we are
concerned that in the system boundaries are not broad enough, the
tolerability and nature of the risks are not sufficiently elaborated,
and that the assurance methods lack theories that allow behaviours to
be assured with adequate confidence.  We advocate the use of assurance
cases based on Assurance 2.0 to support decision making in which the
criticality of the decision as well as the criticality of the system
is evaluated.  Finally, we point out the orders of magnitude
difference in confidence needed in critical rather than everyday
systems, and how everyday techniques do not scale in rigour.

We mapped our findings in some detail to two of the questions posed
by the FAISC organisers ``How can we adapt best practices from other
industries?'', and ``How do they (AI safety frameworks) need to change
when AI systems become capable of posing intolerable levels of risk?''

For the question ``What are common challenges for companies that are yet
to produce a frontier AI system and/or a safety framework?'' we suggest
a significant challenge will be implicit pressure to follow the pattern
of existing frameworks rather than consider the issues independently.

Finally, in response to the questions ``What kinds of resources would
they find helpful?'' and ``How can governments, academia, companies
and civil society, and other third parties support them better?''  we
note that the engineering of critical systems has evolved through open
and diverse discussion of relevant topics and this should also be
encouraged and supported for frontier AI, building on the dialogues at
FAISC.

\subsection*{Acknowledgements}\label{acknowledgements}

Robin Bloomfield's work was part funded by City and St George's,
University of London, and John Rushby's work was funded by SRI
International.

\newcommand{\doi}[1]{\href{https://doi.org/#1}{\tt DOI:#1}}
\addcontentsline{toc}{section}{References}
\bibliographystyle{modplain}

\appendix

\section*{Appendix: Assurance 2.0 in a Nutshell}
\addcontentsline{toc}{section}{Appendix: Assurance 2.0 in a Nutshell}
\label{appendix}

The next two pages reproduce a succinct overview of Assurance 2.0.
Additional material on Assurance 2.0 can be found at
\url{http://www.csl.sri.com/users/rushby/assurance2.0}.



\section*{Supplementary material}

# Assurance 2.0 in a Nutshell

Robin Bloomfield (City, Univ. of London) and John Rushby (SRI)

SRI CSL Technical Note, 14 October 2024

This is intended as a memory aid, not a replacement for reading the longer documents that can be found (as can this) at <https://www.csl.sri.com/users/rushby/assurance2.0>.

**Purpose of Assurance 2.0**: it's a rigorous and systematic approach to developing, presenting, and examining assurance cases to support *indefeasible confidence* in safety or other critical properties

- **Structure**: Claims, Argument, Evidence (CAE), plus Theories and Defeaters
    - **Claims**: precise and meaningful statements about system and environment, presented as atomic propositions in natural language. Some may be marked as **assumptions**
        * Claims may state probabilistic properties and uncertainties (e.g., `pfd` $< 10^{-4}$)
    - **Argument**: typically presented as a tree-like structure of nodes; each node has a **parent** claim, one or more **subclaims**, and usually a **side-claim**
        * Just 5 kinds of (building) **blocks** for argument nodes: **concretion**, **substitution**, **decomposition**, **calculation**, **evidence incorporation**. See Figure 1
        * **Conjunction** of subclaims and side-claim should *deductively* entail parent claim; otherwise flag as *inductive* & apply special care such as confirmation theory (below)
        * **Disjunctive** decompositions are available (useful in *refutational* subcases, see over)
        * **Side-claim** typically factors out deductiveness conditions (e.g., subclaims partition parent claim, or parent claim distributes over components enumerated in subclaims)
        * A **narrative justification**...*justifies* all this; may cite an external **theory**
        * LLMs can interpret claims as *knowledge graphs* over standardized ontology, which can then be checked for consistency using *answer set programming* [1]
    - **Evidence**: a coherent *assembly* of reviews, analyses, tests etc. that *measures* some property of the system. The measurement in turn supports some *useful* inference. This is justified by a **narrative description** that may cite an external **theory**
        * Parent claim of an evidence incorporation block is called the **measured claim**: it says what the evidence is (e.g., testing achieved MC/DC coverage with no faults)
        * Above that is a substitution block that derives a **useful claim** from the measured claim; it says what the evidence means (e.g., there is no unreachable code)
        * Weight of evidential support for the useful claim is examined using the measures of **confirmation theory**, e.g., (Keynes): $\log \frac{P(C \mid E)}{P(C)}$, or (Good): $\log \frac{P(E \mid C)}{P(E \mid \neg C)}$

- **Theories** are self-contained technical descriptions and assurance arguments for specific assurance methods (e.g., static analysis) or (sub)systems (e.g., altitude hold). They include narrative justifications for their arguments and may serve as **templates** for assurance (sub)cases

    - Subcases can be instantiations of parameterized (and ideally *pre-certified*) theories
    - Instantiations can be *expanded* in place (like a macro), or *referenced* (like a subroutine)
    - Much of a case can be *synthesized* from a library of such parameterized theories
    - Standards bodies should deliver theories not guidelines.
    - Overall case can be summarized by enumerating its theories

- **Defeaters** are used to challenge a case, have their own subcases to *refute* or *support* them
  - **Exact Defeaters** introduce negation & refutation: support *eliminative argumentation*
  - Other kind are called **exploratory defeaters** and must eventually be *refuted* (but can then be retained as commentary), or *accepted* as **residual risks**

- An **assurance case** is a package of claims, argument, evidence, plus all supporting theories and narratives; deployment **decision** may be justified in a **sentencing statement**
  - The argument must be **completed**: a *connected* tree/graph where leaves are either evidence, assumptions, or residual risks (or references to completed subcases)
  - Must have no unrefuted defeaters, except those identified as **residual risks**

- **Assessment** employs 4 perspectives: logical, probabilistic, dialectical, and residual risks
  - **Logical assessment** requires a completed argument that is logically valid and **indefeasibly sound**: no credible new information would change the judgement
  - Also, there are (fairly weak) ways to externally assess **probabilistic confidence** in a case. Main value is supporting principled ways of graduating effort vs. risk.
  - **Dialectical examination** combats complacency and confirmation bias: uses defeaters (for claims and argument nodes) and confirmation measures (for evidence).
  - **Residual doubts** are assessed for quantity & risk and all but negligible risks eliminated

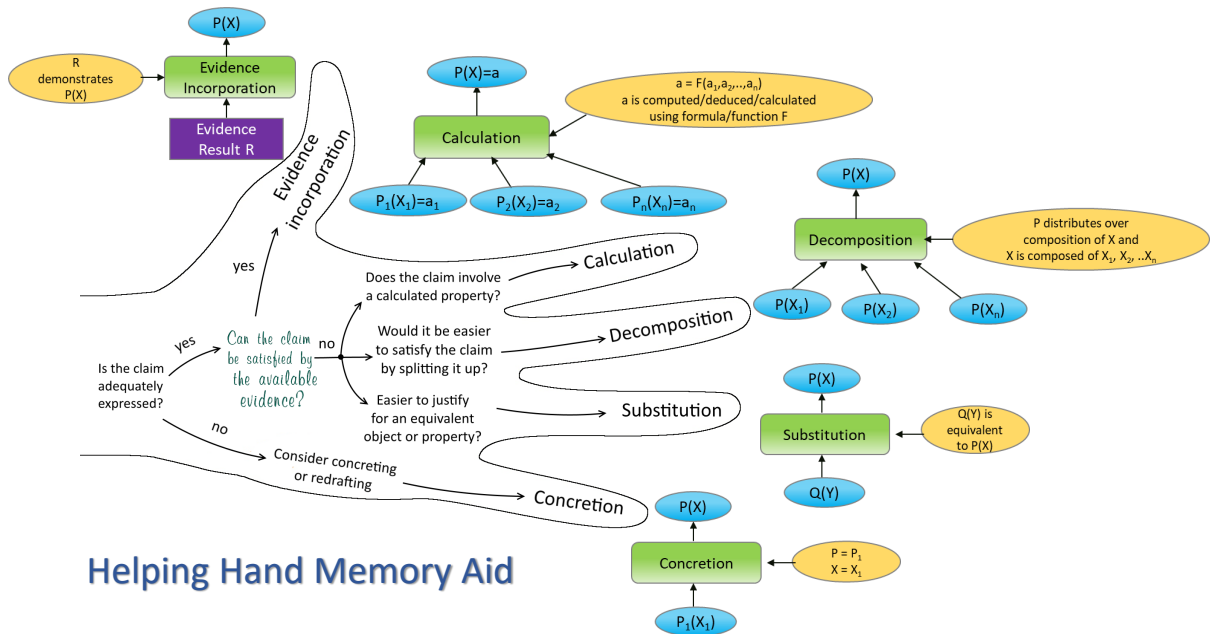

Figure 1: Assurance 2.0 Building Blocks and "Helping Hand" Mnemonic (from [2])



\end{document}